\documentclass[letter]{aa} 

\newcommand{\kms}{\hbox{km\,s$^{-1}$}}
\newcommand{\msun}{$M_{\odot}$}
\newcommand{\teff}{$T_{\rm eff}$}
\newcommand{\lsiv}{\object{LS\,IV$-$14$\degr$116}}
\newcommand{\feige}{\object{Feige\,46}}

\usepackage{natbib}
\usepackage{graphicx}
\usepackage{txfonts}

\begin{document}

\title{Discovery of a second pulsating intermediate helium-enriched sdOB star}

\author{ M. Latour\inst{1}, E. M. Green\inst{2} and G. Fontaine\inst{3} }  

\institute{
Institute for Astrophysics, Georg-August-University, 
Friedrich-Hund-Platz 1, 37077 G\"{o}ttingen, Germany
  \email{marilyn.latour@uni-goettingen.de } 
\and
Steward Observatory, University of Arizona, 
933 North Cherry Avenue, Tucson, AZ 85721, USA
\email{bgreen@as.arizona.edu} 
\and
 D\'epartement de Physique, Universit\'e
  de Montr\'eal, Succ. Centre-Ville, C.P. 6128, Montr\'eal, QC, H3C 3J7, Canada 
  \email{fontaine@astro.umontreal.ca } 
  }

\date{Received ; accepted }

\abstract{
We present the discovery of long-period, low-amplitude, $g$-mode pulsations in the intermediate He-rich hot subdwarf (sdOB) star Feige 46. Up until now only one other He-enriched sdOB star (\lsiv) was known to exhibit such pulsations. From our ground-based light curves of Feige 46, we extracted five independent periodicities ranging from 2294 s to 3400 s.
We fitted our low-resolution, high signal-to-noise optical spectrum of the star with our grid of non-LTE model atmospheres and derived the following atmospheric parameters: \teff\ = 36120 $\pm$ 230 K, log $g$ = 5.93 $\pm$ 0.04 and log $N$(He)/$N$(H) = $-$0.32 $\pm$ 0.03 (formal fitting errors only). These parameters are very similar to those of \lsiv\ and place Feige 46 well outside of the instability strip where the hydrogen-rich $g$-mode sdB pulsators are found. 
We used the Gaia parallax and proper motion of Feige 46 to perform a kinematic analysis of this star and found that it likely belongs to the Galactic halo population. This is most certainly an intriguing and interesting result given that \lsiv\ is also a halo object.
The mechanism responsible for the pulsations in these two peculiar objects remains unclear but a possible scenario involves the $\epsilon$-mechanism. 
Although they are the only two members in their class of variable stars, these pulsators appear to have more in common than just their pulsation properties.  
}


 \authorrunning{Latour, Green \& Fontaine}
 \titlerunning{Discovery of a second intermediate He-rich pulsating sdOB star}

\maketitle

\section{Introduction}

The vast majority of subdwarf B (sdB) stars\footnote{We note that the term "sdB stars" is often understood to include the slightly hotter sdOB stars, since they are nearly identical in an evolutionary sense.} are compact, hot, helium-core burning objects that lost almost all of their hydrogen envelope prior to the He-flash, leaving them unable to sustain H-shell burning. Consequently they have masses close to the canonical value required for the He-flash ($\sim$0.48 \msun), to which their remaining hydrogen envelope contributes minimally ($\la$ 0.02 \msun). 
The discovery of the first pulsating sdB stars \citep{kil97} two decades ago proved to be a stepping stone for our understanding of these particular objects (see \citealt{heb16} and \citealt{cha16} for recent reviews). Their asteroseismic modeling has allowed, among other things, the probing of their structural properties such as core size and composition (e.g.\ \citealt{cha11,groo13}), rotation \citep{cha18} and stellar masses (\citealt{font12} and references therein). Pulsating hot subdwarfs show pressure ($p$-) and gravity ($g$-) mode instabilities. The former appear as rapid periodic variations of a few millimag on a timescale of a few minutes, while the latter are characterized by slower variations with periods between 0.5 and 4 hrs and even smaller amplitudes, $\la$ 1 millimag. The two types of sdB pulsation modes are excited in two distinct instability regions in the log $g$ - \teff\ plane, with the slow, $g$-mode pulsators (formally referred to as {\it V1093 Her} stars) found at cooler effective temperatures (22$-$29 kK) than their rapidly pulsating {\it V361 Hya} counterparts (29$-$36 kK). Interestingly, some stars at the boundary between the two instability domains show both $p$- and $g$-mode pulsations (e.g.\ \citealt{schuh06}). The instability regions are well defined observationally and theoretically \citep{cha01,bloe14} although the boundary between the two types becomes more fuzzy when viewed with space-borne sensitivity. Both classes of pulsating hot subdwarfs can be successfully modeled in terms of the same driving mechanism: a classical opacity ($\kappa$) mechanism associated with a local overabundance of iron and nickel in the driving region \citep{cha97}. 

While the majority of pulsating hot subdwarfs belong to these two classes, each well described by asteroseismology, a few of these pulsators remain theoretically challenging: the few hotter sdO pulsators \citep{ran16,kil17} and the very unusual sdOB\footnote{Although \lsiv\ has been previously referred to as an sdB star, its spectrum, showing a strong \ion{He}{ii} 4686 \AA\ line, is better classified as  sdOB, a more specific term for objects in the transition region between sdB and sdO stars.}
star \lsiv\ \citep{green11}. The latter shows long-period ($P$ $\sim$ 2000$-$5000 s), multiperiodic luminosity variations in spite of the fact that its atmospheric parameters place the star on the hotter side of the $p$-mode instability strip where its long pulsation periods cannot be explained by the iron-bump $\kappa$-mechanism \citep{ahm05,green11}. In addition to its distinctive pulsation properties, \lsiv\ has very peculiar atmospheric abundances: it contains the largest overabundances of Sr, Y, and Zr reported in a hot subdwarf as well as being one of the very few known to have an atmosphere enriched in helium \citep{nas11}. In addition, \citet{ran15} found that it belongs to the halo population based on its kinematic properties. The star was therefore suggested to be the prototype of a new class of pulsating hot subdwarfs referred to as He-sdBV \citep{kil10}. However, \lsiv\ remained the sole member of its class for almost a decade\footnote{
Two other variable helium-enriched hot subdwarfs were discovered with the {\it Kepler} satellite (UVO 0825+15 and KIC 1718290; \citealt{jeff17,ost12}), but given their discrepant pulsation properties and, in the case of KIC 1718290, lower \teff, they likely do not belong to the same class of pulsators as \lsiv.}. 

In this Letter, we present \feige\ as the long sought-after second member of the He-sdBV class. Our extensive photometric monitoring of the star has revealed the presence of multiperiodic variations very similar to those seen in \lsiv\ and our spectral analysis demonstrates that the atmospheric parameters of both stars are strikingly similar. This discovery is all the more exciting given that \feige\ is bright enough ($V$ = 13.0) to be observed by the TESS satellite and to allow high-resolution spectroscopic follow-up and more detailed investigation of its atmospheric properties.
 
\section{Pulsational properties}\label{sec:ident}

\begin{table}
\caption{Journal of Photometric Observations for Feige 46}\label{obs}
\centering
\small
\begin{tabular}{c c c c}
\hline
\hline
Date & Start of Run & Number of Frames & Length \\
(UT) & (HJD2458170+) & (h)\\
\hline 
2018-Feb-26 & 5.87597 & 425 & 3.613 \\
2018-Feb-27 & 6.72110 & 853 & 7.251 \\
2018-Mar-06 & 13.71209 & 556 & 4.726 \\
2018-Mar-25 & 32.66519 & 300 & 2.550 \\
2018-Mar-27 & 34.76409 & 498 & 4.233 \\
2018-Mar-28 & 35.64764 & 850 & 7.225 \\
2018-Mar-29 & 36.71110 & 286 & 2.431 \\
2018-Apr-02 & 40.61363 & 906 & 7.701 \\
2018-Apr-03 & 41.61282 & 911 & 7.744 \\
2018-Apr-14 & 52.61746 & 771 & 6.554 \\
2018-Apr-15 & 53.61904 & 768 & 6.528 \\
2018-Apr-29 & 67.23060 & 563 & 4.786 \\
2018-Apr-30 & 68.62797 & 271 & 2.304 \\
2018-May-25 & 93.63795 & 441 & 3.749 \\
\hline
\end{tabular}
\end{table}

\begin{table*}[h]
\centering
\caption{Harmonic Oscillations Detected in the Light Curve of Feige
46}\label{period}
\centering
\begin{tabular}{c c c c c}
\hline
\hline
Period & Frequency & Amplitude & Phase & (S/N) \\
(s) & ($\mu$Hz) & (\%) & (s)   &  \\
\hline 
2294.673 $\pm$ 0.020\tablefootmark{a} & 435.792 $\pm$ 0.004 & 0.290 $\pm$ 0.015 & 
420 $\pm$ 19 & 15.1 \\
2296.064 $\pm$ 0.024\tablefootmark{a} & 435.528 $\pm$ 0.005 & 0.243 $\pm$ 0.015 & 
1056 $\pm$ 23 & 12.7 \\
2585.797 $\pm$ 0.083 & 386.728 $\pm$ 0.012 & 0.090 $\pm$ 0.015 &
670 $\pm$ 70 & 4.7 \\
2757.738 $\pm$ 0.032 & 362.616 $\pm$ 0.004 & 0.267 $\pm$ 0.015 &
2210 $\pm$ 25 & 13.9 \\
2999.184 $\pm$ 0.072 & 333.424 $\pm$ 0.008 & 0.140 $\pm$ 0.015 &
1528 $\pm$ 53 & 7.3 \\
3400.713 $\pm$ 0.057 & 294.056 $\pm$ 0.005 & 0.225 $\pm$ 0.015 &
2520 $\pm$ 37 & 11.7 \\
\hline
\end{tabular}
\centering
\tablefoot{
\tablefoottext{a}{These modes are part of a possible rotation triplet}
}
\end{table*} 

\begin{figure*}
 	\centering
    \includegraphics[width=0.49\textwidth]{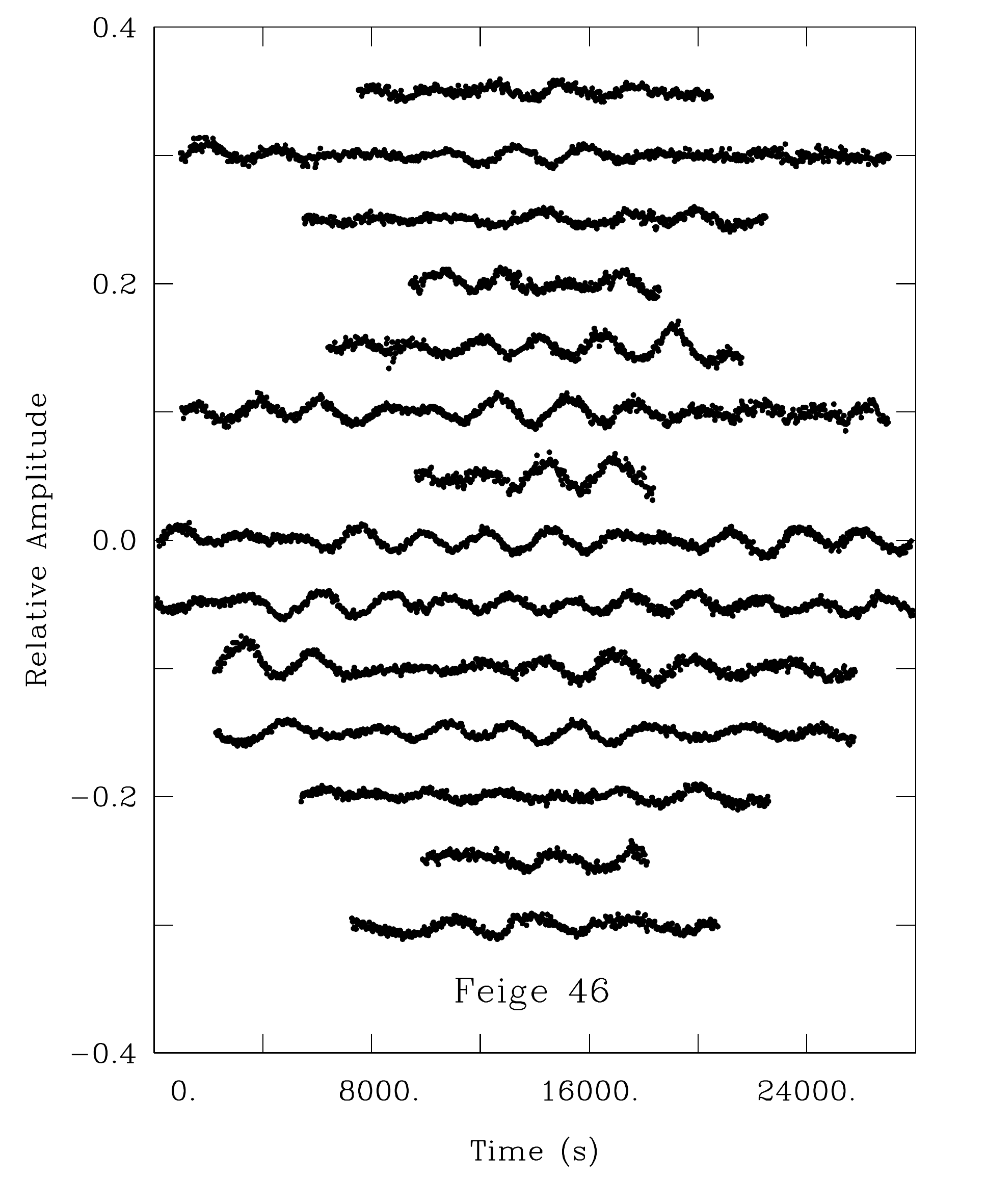}
    \includegraphics[width=0.49\textwidth]{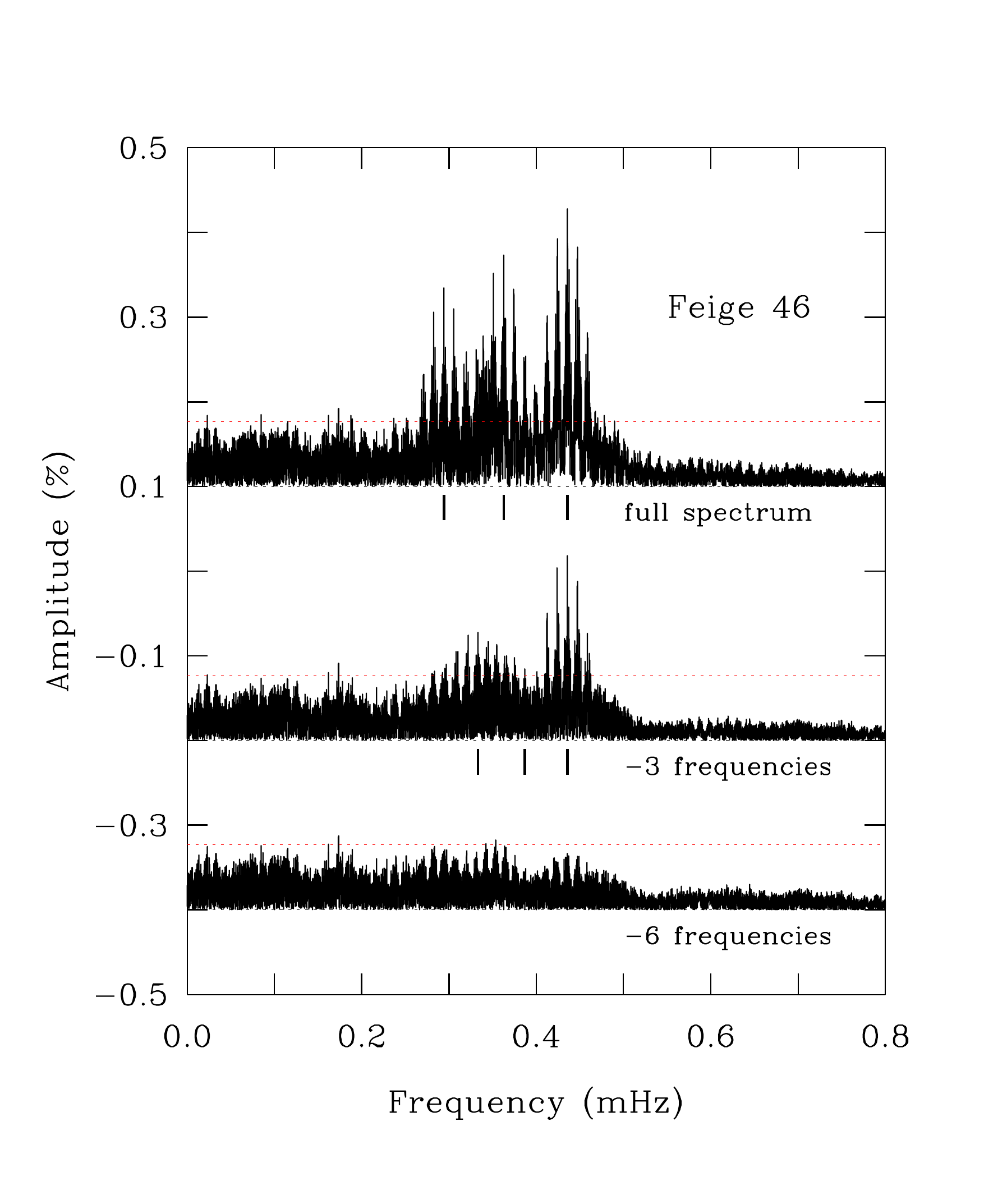}
    \caption{Left - All light curves obtained for \feige. The data have been shifted arbitrarily along the x- and y-axes for visualization purposes. The y-axis is expressed in units of fractional brightness intensity (the residual amplitude of the pulsation relative to the mean brightness of the star). 
     Right - Zoomed in view of the Fourier transform of the entire data set in the 0$-$0.8 mHz range where the periodicities are found. The lower transforms show the successive steps of prewhitening by the three and six frequencies indicated. The dashed horizontal lines indicate the 4$\sigma$ noise level. }
             \label{ft}
\end{figure*}

The variable nature of \feige\ was discovered in February 2018 during a photometric search for asteroseismically interesting hot subdwarf stars at the Steward Observatory 1.55 m Kuiper telescope on Mt.\ Bigelow, using the Mont4K CCD camera\footnote{http://james.as.arizona.edu/~psmith/61inch/instruments.html}. Following the realization that this star closely resembled \lsiv, with low-amplitude luminosity variations on a timescale of about 45 minutes at odds with its sdOB spectrum (see Section 3), extensive follow-up was performed during 13 additional nights on the Kuiper telescope (see Table \ref{obs}). This resulted in 71.19 hrs of time-series observations with an average sampling time of 30 s, taken through a broadband Schott 8612 filter. The images were reduced using standard IRAF photometric data reduction tasks. The light curves were constructed relative to five reasonably bright reference stars distributed symmetrically around \feige\ in the 9.7\arcmin\ x 9.7\arcmin\ Mont4k field of view. The photometric aperture size was set to either 2.25 times the average FWHM of the six measured stars or 15 pixels (= 6.45\arcsec), whichever was smaller, to avoid contamination from the fainter visual companion 13.3\arcsec\ northwest of \feige. Light curves for each of the reference stars relative to the four others confirmed that none were variable above the level of the photometric noise.  The light curves for \feige\ are shown in the left panel of Fig. \ref{ft}, in the same order as the observations in Table \ref{obs}.

 \begin{figure*}
 	\centering
 	\includegraphics[angle=270, width=0.75\textwidth]{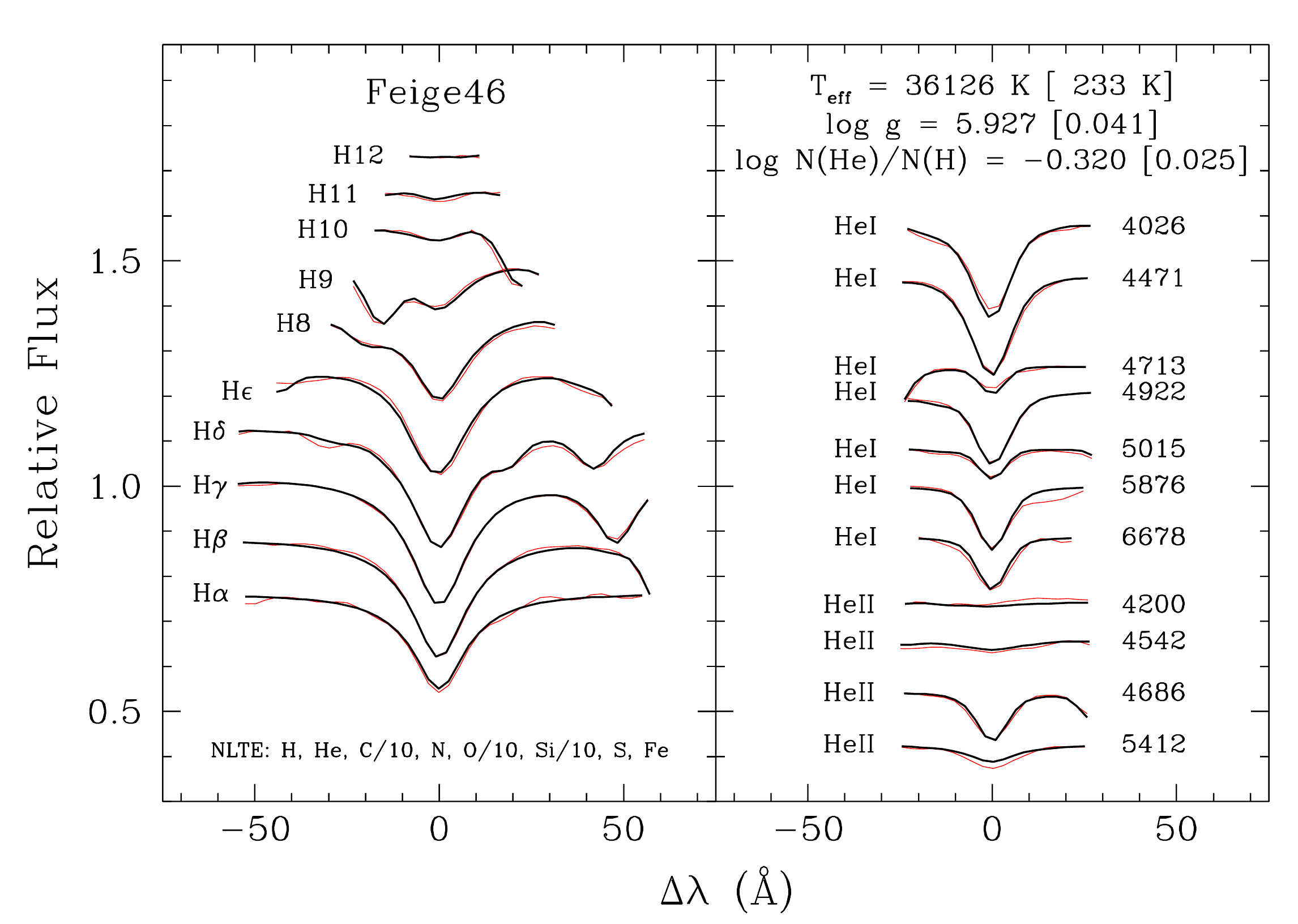}
	\caption{Best fit (black curve) to the Balmer and helium lines visible in our high $S/N$, low resolution spectrum (red curve) of \feige. We used a three-dimensional grid (\teff, log~$g$, $N$(He)/$N$(H)) of NLTE model atmospheres including a fixed metallic composition (solar N, S, and Fe, 1/10 solar C, O, and Si) typical for sdB stars.}
 	\label{sfit}
 \end{figure*}

The time-series photometry gathered for \feige\ was analyzed in a standard way using a combination of Fourier analysis, least-square fits to the light curve, and pre-whitening techniques, as described in \citet{bill00}. The light curve analysis revealed the presence of six frequency peaks whose characteristics are listed in Table \ref{period}. The phases given in the table are relative to the beginning of the first run on UT 2018 Feb 26, The uncertainties and S/N are computed as described in \citet{green11}. We found that \feige\ has oscillation periods in the 2294$-$3400 s range, which overlaps quite well with the range of periods detected in \lsiv\ (1953$-$5083 s; \citealt{green11}). The low amplitudes are typical of $g$-mode pulsations. 
The right panel of Fig.\ \ref{ft} presents the Fourier amplitude spectrum of the full data set (upper curve) in the 0$-$0.8 mHz range where the pulsating periods of \feige\ are found. The lower curves are the resulting Fourier transforms after prewhitening by the three and six frequency peaks indicated. The main oscillation peak revealed a structure compatible with the rotational splitting of an $l$=1 mode, with two out of the three components having a S/N above the 4.5$\sigma$ limit\footnote{A central component is present at a frequency of 435.670 $\pm$ 0.014 $\mu$Hz, but only with S/N = 3.4.}. The average frequency difference ($\Delta f$) between the peaks is 0.132 $\mu$Hz. Assuming these are high-radial order $k$ modes, this frequency splitting is indicative of a slow rotation with a period of $\sim$43 days.  

 \section{Spectroscopic properties}

\feige\ is part of the Arizona-Montr\'{e}al spectroscopic sample \citep{font14} and was observed multiple times at Steward Observatory's 2.3 m Bok Telescope on Kitt Peak between 2004 and 2011. The observations were obtained with the B\&C spectrograph using the 400 mm$^{-1}$ grating in first order and a 2.5\arcsec\ slit,  resulting in a wide wavelength coverage (3620$-$6900 \AA) and a rather low resolution of 8.7 \AA. Our final high-S/N spectrum was constructed from five median-filtered individual spectra which were first cross-correlated to correct for possible variations in radial velocity (RV). However, no RV variations were found above the observational noise level ($\sigma \sim$10~\kms).

We derived the atmospheric parameters (\teff, log $g$, and helium abundance) of \feige\ using our grid of non-LTE model atmospheres, which are especially well suited for analysis of sdB stars. The models and synthetic spectra were computed using the TLUSTY and SYNSPEC codes \citep{hub11}, including a metallic composition typical of sdB stars \citep{bla08,bra10}. Figure \ref{sfit} shows the resulting best fit to the normalized Balmer and helium lines present in the observed spectrum. We find \teff\ = 36100 $\pm$ 230~K, log $g$ = 5.93 $\pm$ 0.04, and a helium abundance of log $N$(He)/$N$(H) = $-$0.32 $\pm$ 0.03. The quoted uncertainties are formal errors returned by the $\chi^2$-minimization procedure and are therefore only lower limits. 
Since the same model atmosphere grid was used and the spectra were obtained with the identical instrumental setup, the atmospheric parameters of \feige\ are directly comparable to those derived for \lsiv\ by \citet{green11}, \teff\ = 34900~K, log $g$ = 5.9 and log $N$(He)/$N$(H) = $-$0.6\footnote{The analysis of a higher resolution spectrum of \lsiv\ with the same models corroborate these values \citep{ran15}.}. Thus both stars have similar atmospheric parameters, in addition to comparable luminosity variations.  Although \feige\ is somewhat 
more helium-enriched than \lsiv, it is still classified as an intermediate He-rich subdwarf, since its helium abundance is above solar (log $N$(He)/$N$(H) = $-$1), yet its atmosphere is still dominated, although only slightly, by hydrogen. 
The slightly higher \teff\ and stronger \ion{He}{ii} 4686 line of \feige\ with respect to \lsiv\ likely explain why the former was classified early on as an sdO star \citep{feige58,grah70}, while the latter was considered an sdB. The discrepancy in spectral type between the two stars is however at odds with their similar atmospheric parameters. In fact, both stars are more accurately described as sdOB given that \ion{He}{i} and \ion{He}{ii} lines are clearly visible in their spectra. Therefore, we now refer to the two pulsators as He-sdOBV instead of using the previous He-sdBV nomenclature. This allows them to be distinguished from the cooler He-sdBs (such as JL 87 and KIC 1718290) as well as the hotter He-sdOs, which have different properties than those of \lsiv\ and \feige\ (see also section 2.2 of \citealt{heb09} for additional details on spectral classification). On the other hand, given the variety of spectral classification scheme and the growing number of pulsation properties being discovered in hot subdwarfs, it might be clearer to refer to these two particular pulsators, and potential future members of their class, by the name of the prototype: \lsiv, also named \textit{V366 Aqr} in the general catalogue of variable stars \citep{var_cat17}. 

\feige\ has been little studied in the past. The most recent atmospheric analysis we found in the literature is that of \citet{bauer95}, who derived \teff\ = 37500 $\pm$ 1500~K, log~$g$ = 5.25 $\pm$ 0.25, and abundances of C and N close to solar. They also estimated a spectroscopic distance of 1200$^{+535}_{-370}$ pc. Their analysis was based on data obtained with the CASPEC Cassegrain echelle spectrograph mounted on the 3.6 m ESO telescope at La Silla Observatory in Chile. On the other hand \citet{kud76} derived \teff\ $\sim$ 41000~K and log $g$ $\sim$6.3 based on Str\"{o}mgren and Balmer jump colors.  

Using newly available parallax and photometric measurements from Gaia, the derived atmospheric parameters can be verified by computing the spectroscopic distance and comparing with the distance derived from the parallax. 
We performed this exercise for both \lsiv\ and \feige\, assuming a mass of 0.47 \msun, consistent with the mean sdB mass reported by \citet{font12}, and using our model spectra to compute synthetic G(BP) and G(RP) absolute magnitudes. Comparison with the apparent magnitudes from Gaia DR2 resulted in spectroscopic distances of 426 $\pm$ 27 pc for \lsiv\ and 514 $\pm$ 31 pc for \feige. These agree very well with the distances derived from their Gaia parallaxes, 420$^{+15}_{-14}$ pc and 538$^{+20}_{-19}$ pc, respectively, overlapping within 1$\sigma$. In addition, the inferred reddenings, E(B$-$V) = 0.028 $\pm$ 0.006 for \lsiv\ and 0.005 $\pm$ 0.005 for \feige, are smaller than the maximum values of 0.038 and 0.028 along the two lines of sight determined by \citet{schla11}.

\section{Discussion}

Intermediate He-rich subdwarfs are relatively rare objects \citep{geier17} and some show extreme overabundances of trans-iron elements (Ge, Sr, Y, Zr, and Pb; \citealt{jeff17,wild18}).  Among these, \lsiv\ stands out as having an atmosphere especially enriched in Sr, Y, and Zr \citep{nas13}. Such large overabundances render these elements detectable not only in their UV spectra, as commonly seen in other sdBs, but also in the optical region. 
It seems plausible that the atmosphere of \feige\ may also display similar chemical peculiarities. 
\citet{otoole04} reported the detection of Pb in a spectrum of \feige\ observed with the Goddard High Resolution Spectrograph (GHRS) on board HST ($\sim$1320$-$1520 \AA), but there has been no modern
high-resolution optical spectroscopy for this star.  Further analysis of the GHRS data, especially in combination with new optical data, would be of utmost interest to address the chemical portrait of \feige. 

\begin{figure}
 	\centering
 	\includegraphics[width=0.48\textwidth]{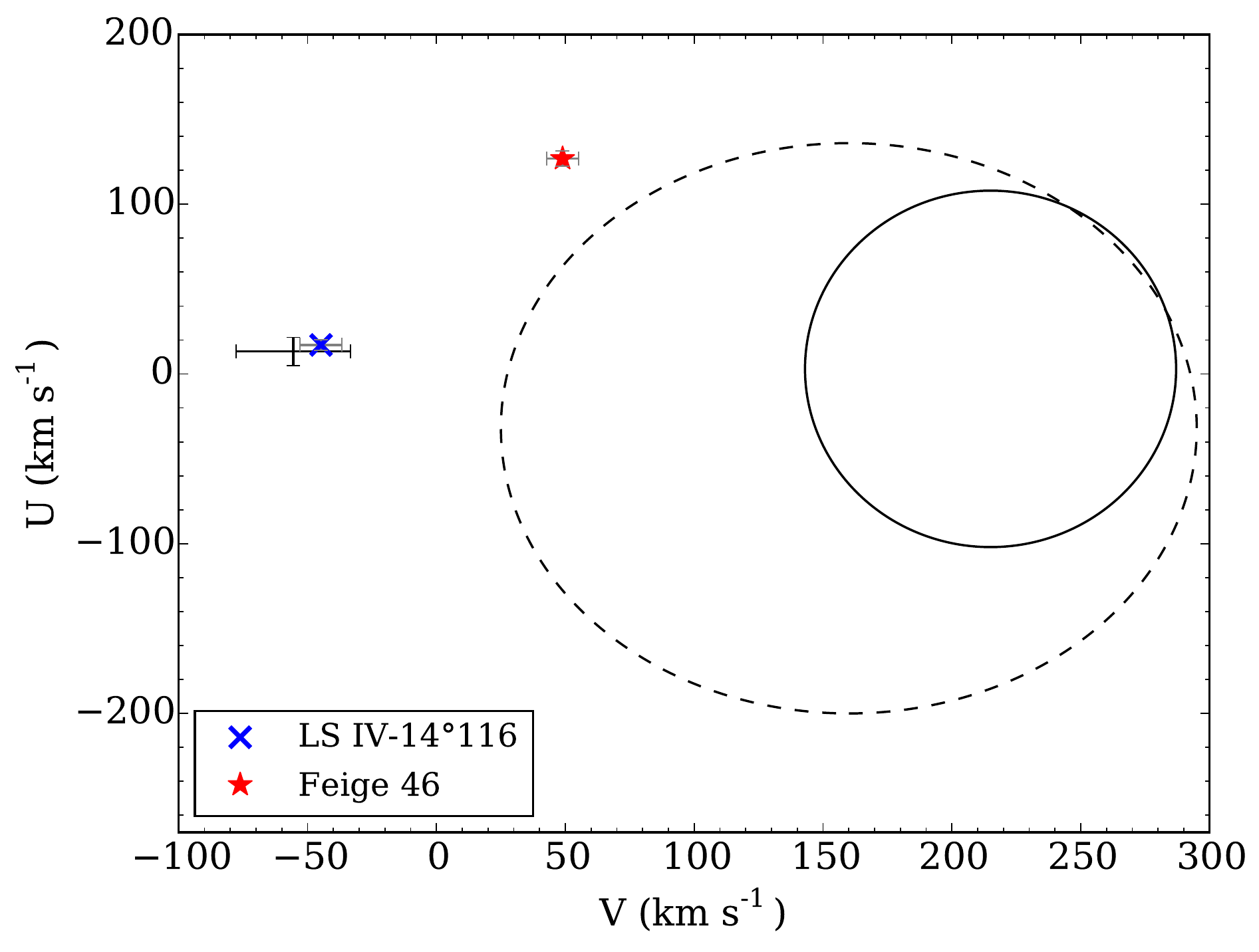}
	\caption{$U-V$ diagram showing the positions of \feige\ (star) and \lsiv\ (cross) compared to the 3$\sigma$ thin-disk (solid line) and thick-disk (dashed line) contours as defined by \citet{pauli06}. Stars outside these contours have kinematic properties of the halo population. The larger error bars are centered on the values derived by \citet{ran15} for \lsiv.}
 	\label{kinem}
 \end{figure}

Another peculiar aspect of \lsiv\ is its membership in the Galactic halo population, which was confirmed by its Galactic velocity components and retrograde orbit around the Galactic center \citep{ran15}. We conducted a similar analysis using the Gaia DR2 proper motion and parallax data \citep{gaiadr2}.
We used the radial velocity value reported by \citet{dri87} of 90~\kms\ with a more conservative error of 4~\kms.  We then followed the procedure described by \cite{ran15} to derive the Galactic radial and rotational velocity components of \feige\ ($U$ = 126.9 $\pm$ 4.5 \kms, $V$ = 48.9 $\pm$ 6.1 \kms), and to re-compute those of \lsiv\ ($U$ = 17.1 $\pm$ 3.1 \kms, $V$ = $-$44.8 $\pm$ 7.9 \kms). The positions of the two stars in the $U-V$ diagram are shown in Fig.~\ref{kinem} along with the 3$\sigma$ contours expected for the thin and thick disk populations. Like \lsiv, \feige\ lies outside these contours, indicating that it has kinematic properties typical of the halo population. Still, halo kinematics do not necessarily imply that the stars were formed in situ. Some sdBs with extreme kinematics are likely to have acquired such properties after dynamical interactions, e.g. with the central black hole of the Milky Way, \citep{til11,zieg17}. Nevertheless, the kinematics of \lsiv\ and \feige\ are not particularly extreme and a halo origin remains a plausible explanation.

The similarity between \feige\ and \lsiv\ in terms of both photometric variability and atmospheric properties leaves little doubt that the two stars are pulsating via the same mechanism. 
Their variability cannot be explained by the iron-bump $\kappa$-mechanism at work in the vast majority of pulsating hot subdwarfs. Their temperatures and surface gravities place them in the instability region where only $p$-modes would be excited by this mechanism, yet such short-period pulsations are not observed in either star, only the unanticipated long-period ones. The possibility that a strong magnetic field produces the chemical anomalies and luminosity variability seen in \lsiv\ was suggested by \citet{nas11}, but later discarded based on spectropolarimetric observations of the star \citep{ran15}. The radial velocity variations seen in time-resolved spectroscopy suggest instead that the pulsations are associated with an oscillatory motion of the stellar surface \citep{jeff15,mar17}. 

Currently, the most promising explanation for the variability of these stars is pulsations driven by nuclear burning ($\epsilon$-mechanism) in helium subflashes following a delayed core He-flash \citep{bert11}. 
A delayed core flash occurs after the star has evolved away from the red giant branch and can happen as late as when the star is already contracting on the white dwarf cooling curve \citep{cas93,lanz04}. The delayed flash is more explosive and causes extra mixing between the helium-rich material in the core and the hydrogen-rich superficial layers, producing an atmosphere enriched in helium. This late-flasher scenario is therefore an evolutionary channel invoked to explain the formation of helium-enriched hot subdwarfs \citep{bert08}. 
In their exploration of the pulsation properties of late-flasher evolutionary models, \citet{bat18} showed that long-period $g$-modes can be excited via the $\epsilon$-mechanism in the log $g$ $-$ \teff\ region where both pulsators are found. Although the predicted modes have periods shorter than 2000 s, below the observed period range, \citet{bat18} pointed out that this could be due to shortcomings in the evolutionary models, since the details of this short evolutionary phase are poorly constrained. Interestingly, pulsations excited by the $\epsilon$-mechanism have not yet been confirmed in any type of star \citep{sow18}. These two He-sdOBVs are currently the most promising candidates for this pulsation mechanism.

An alternative $\kappa$-mechanism was recently suggested by \citet{saio19}, in which the excitation is driven by carbon and oxygen opacity bumps. Although their models can excite pulsations in the observed period range, the stellar structure needed to do so, a 0.5 \msun\ helium main-sequence star having a substantial enrichment of carbon and/or oxygen in the envelope, is very challenging to explain from an evolutionary perspective.

\section{Conclusion}

We report the discovery of long-period $g$-mode pulsations in the intermediate He-rich sdOB star \feige. The atmospheric parameters and helium abundance of the star confirm that this object belongs to the He-sdOBV pulsating class, which was until now defined only by the unique prototype \lsiv.  The excitation mechanism responsible for the observed pulsations in both stars is a matter of debate, with the two suggested models, the $\epsilon$-mechanism and the $\kappa$-mechanism via carbon and oxygen opacity-bumps, each having their own shortcomings. In this context, the pulsation periods observed in the two stars, and potentially in other similar objects, should be of great help in testing and refining evolutionary models. The kinematic properties of \feige\ indicate that it belongs to the Galactic halo population, just like \lsiv.  The halo origin as well as the probable single nature and slow rotation of both stars put further constraints on the evolutionary history of these enigmatic objects. 
\begin{acknowledgements}
 M.L.\ acknowledges funding from the Deutsche Forschungsgemeinschaft (grant DR 281/35-1). This research has made use of NASA's Astrophysics Data System. We would like to thank Ulrich Heber for useful discussions.
 \end{acknowledgements}

\bibliographystyle{aa}

\end{document}